\documentclass[prd,aps,floatfix,preprint,nofootinbib]{revtex4}
\usepackage{amsmath,amssymb,amsbsy,color}
\usepackage{epsfig}

\begin{document}
 \vspace*{-20mm}
 \begin{flushright}
  \normalsize
  KEK-CP-211,\
  UTHEP-561,\
  YITP-08-46
 \end{flushright}
\title{S-parameter and pseudo-Nambu-Goldstone boson mass from lattice
QCD}
\author{
E. Shintani$^{a}$, 
S. Aoki$^{b,c}$,
H. Fukaya$^{a,d}$,
S. Hashimoto$^{a,e}$,
T. Kaneko$^{a,e}$,
H. Matsufuru$^{a}$,
T. Onogi$^{f}$,
N.~Yamada$^{a,e}$\\
(JLQCD Collaboration)
}
\affiliation{
$^a$High Energy Accelerator Research Organization (KEK), Tsukuba
    305-0801,Japan\\
$^b$Graduate School of Pure and Applied Sciences, University of
    Tsukuba, Tsukuba 305-8571, Japan\\
$^c$Riken BNL Research Center, Brookhaven National Laboratory, Upton,
    New York 11973, USA\\
$^d$The$\!$ Niels$\!$ Bohr$\!$ Institute,$\!$ The$\!$ Niels$\!$ Bohr$\!$
    International$\!$ Academy,$\!$
    Blegdamsvej$\!$ 17$\!$ DK-2100$\!$ Copenhagen$\!$ {\O},$\!$
    Denmark$\!\!\!$\\
$^e$School of High Energy Accelerator Science, The Graduate University
    for Advanced Studies (Sokendai), Tsukuba 305-0801, Japan\\
$^f$Yukawa Institute for Theoretical Physics, Kyoto University,
           Kyoto 606-8502, Japan
}
\begin{abstract}
 We present a lattice calculation of $L_{10}$, one of the low energy
 constants in Chiral Perturbation Theory, and the charged-neutral pion
 squared mass splitting, using dynamical overlap fermion.
 Exact chiral symmetry of the overlap fermion allows us to reliably
 extract these quantities from the difference of the vacuum polarization
 functions for vector and axial-vector currents.
 In the context of the technicolor models, these two quantities are read
 as the $S$-parameter and the pseudo-Nambu-Goldstone boson mass
 respectively, and play an important role in discriminating the models
 from others.
 This calculation can serve as a feasibility study of the lattice
 techniques for more general technicolor gauge theories.
\end{abstract}
\pacs{12.38.Gc\vspace{-0ex}}
\maketitle

Spontaneous chiral symmetry breaking (S$\chi$SB) of strongly
interacting gauge theory may provide a natural mechanism for the
electroweak symmetry breaking.
A class of new physics models based on this idea, so-called the
technicolor models, has been studied
extensively~\cite{Weinberg:1975gm}.
In most of those models, massless techni-quarks with weak charge are
introduced; the weak gauge bosons acquire masses from their S$\chi$SB.
The $S$-parameter may then be sizably affected, for which those models
can be strongly constrained through the electroweak precision
measurements~\cite{Peskin:1990zt}.
Another characteristic signal of the technicolor models, that may be
observed at the LHC experiments, is the presence of extra
Nambu-Goldstone bosons (NGBs) which are not eaten by the weak gauge
bosons.
They are called the pseudo-NGBs (pNGBs), since they must be made massive
by introducing explicit breaking of the chiral symmetry of the
techni-quarks in a model dependent way, otherwise they would remain
massless.
Since the $S$-parameter and the pNGB mass are consequences of
strong dynamics of the underlying theory, non-perturbative framework
is required for their calculation.
In previous studies, some model was involved in the calculation, 
{\it e.g.} \cite{Harada:2004qn}.

In this work we consider two-flavor QCD as a testing ground of our
method and demonstrate that the first principles calculation of those
quantities are possible.
In this context, the $S$-parameter corresponds to $L_{10}^r$ (or $l_5^r$
in another convention), one of the low-energy constants of the chiral
perturbation theory (ChPT), as
$S$=$-16\pi[L_{10}^r(\mu) - \{ \ln(\mu^2/m_H^2) -1/6\}/192\pi^2]$ with a
renormalization scale $\mu$ and the Higgs mass
$m_H$~\cite{Peskin:1990zt}.
$L_{10}^r$ is related to a difference of vacuum polarization
functions between vector and axial-vector currents
$\Pi_{V-A}^{(1)}(q^2)\equiv\Pi_V^{(1)}(q^2)-\Pi_A^{(1)}(q^2)$
near the zero momentum insertion.
(A formula will be given in (\ref{eq:chpt_v-a_1}).)

For the pNGB mass, a mass formula that is valid for a wide
range of technicolor models and breaking patterns is
known~\cite{Peskin:1980gc}.
The formula contains a nonperturbative part written in terms of the
vacuum polarization functions.
The charged pions in two-flavor QCD is an example of pNGB, as the
electromagnetic interaction explicitly breaks $SU(2)$ chiral symmetry
and gives a finite mass even in the massless limit of up and down
quarks~\cite{Blum:2007cy}.
The corresponding mass formula is known as the
Das-Guralnik-Mathur-Low-Young (DGMLY) sum rule~\cite{Das:1967it}
\begin{equation}
  m_{\pi^\pm}^2 = -\frac{3\alpha}{4\pi}
  \int^{\infty}_0\!\!\!\! dq^2\, \frac{q^2\, \Pi_{V-A}^{(1)}(q^2)|_{m_q=0}}{f^2},
\label{eq:dmpi}
\end{equation}
which gives the mass of charged pions at the leading order of the
electromagnetic interaction.
Here $f$ denotes the pion decay constant in the chiral limit.
Note that neutral pion is massless in this limit.

In the continuum theory chiral symmetry guarantees that the difference
$\Pi_{V-A}^{(1)}(q^2)$ exactly vanishes in the absence of both explicit
and spontaneous chiral symmetry breaking.
Any remaining difference in the absence of explicit breaking thus
signals the S$\chi$SB.
Therefore, the use of exactly chiral fermion formulation is mandatory in
the lattice calculation, in order to avoid fake contributions to
$\Pi_{V-A}^{(1)}(q^2)$ due to non-chiral lattice fermion formulations
such as the Wilson-type fermions.
Here we use the overlap fermion~\cite{Neuberger:1997fp}, which respects
exact chiral symmetry at finite lattice spacings.
Employing this fermion, we have successfully done a precise calculation
of the chiral condensate~\cite{Fukaya:2007fb}, which also requires
excellent chiral symmetry to control systematic errors.

We perform a two-flavor QCD calculation on a $16^3\times 32$ lattice at
a lattice spacing $a$ = 0.118(2) fm determined with the Sommer scale
$r_0$=0.49 fm as an input~\cite{Aoki:2008tq}.
The quark mass in the lattice unit is $\hat m_q$=$am_q$= 0.015, 0.025,
0.035, and 0.050, which roughly cover the range between 1/6 to 1/2 of
the strange quark mass.
The global topological charge $Q$ is fixed to ensure the exact chiral
symmetry~\cite{Aoki:2008tq}.
The main simulations are done in the $Q$=0 sector, using 10,000
trajectories.
At $\hat m_q$ = 0.050, the simulations are also performed in other two
sectors ($Q=-2$ and $-4$) to estimate the finite volume effect due to
fixing $Q$~\cite{Aoki:2007ka}.
For each sea quark mass, the measurements are made at every 50
trajectories.
Statistical errors are estimated from a jackknife analysis with 100
jackknife bins each containing two consecutive measurements.
Details of our configuration generation and the pion spectrum and decay
constant analysis are found in~\cite{Aoki:2008tq} and
\cite{Noaki:2007es}, respectively.

We calculate the current-current correlators for vector and axial-vector
currents to obtain the corresponding vacuum polarization functions.
We use as the vector current
$V^{(12)}_\mu$=$Z\,\bar q_1 \gamma_\mu (1-aD/2m_0)q_2$, where $q_1$ and
$q_2$ represent different flavors of quarks, $D$ the overlap-Dirac
operator in the massless limit, and $m_0$=1.6.
The axial-vector current $A^{(12)}_\mu$ is the same but $\gamma_\mu$
is replaced by $\gamma_\mu\gamma_5$.
The factor $(1-aD/2m_0)$ is necessary
to make the $V$ and $A$ form an exact multiplet under the axial
transformation.
Because of this exact symmetry, then leading to the strong correlation,
even the lattice artifacts and statistical fluctuations cancel between
$VV$ and $AA$ correlators except for the effects of S$\chi$SB.
Indeed, the statistical errors are much smaller than those of the
previous calculations of the $VV$ correlator~\cite{Blum:2002ii}.
The common renormalization constant $Z=1.3842(3)$ is determined
nonperturbatively~\cite{Noaki:2007es}.

Since the continuous rotational symmetry is violated on the lattice at
$O(a^2)$ and the currents we use are not conserved
(cf. \cite{Kikukawa:1998py}), the general form of the current-current
correlator reads
\begin{eqnarray}
    {\Pi_J}_{\mu\nu}(\hat q)
&=& \sum_x e^{i \hat q \cdot x}
    \langle\, 0|\,T\left[J^{(21)}_\mu(x)J^{(12)}_\nu(0)\right]
    |0\,\rangle
\nonumber\\
&=& 
    \sum_{n=0}^\infty B_J^{(n)}(\hat q_\mu)^{2n} \delta_{\mu\nu}
\!+\!\!
    \sum_{n,m=1}^\infty C_J^{(n,m)}(\hat q_\mu)^{2n-1}(\hat q_{\nu})^{2m-1},
\label{eq:j-j_in}
\end{eqnarray}
where $J=V$ or $A$.
$B_J^{(n)}$ and $C_J^{(n,m)}$ are scalar functions of lattice momentum
$\hat q_\mu$=$2 \pi n_\mu/L$ with $n_\mu$ an integer ranging from
$-L/2+1$ to $L/2$ ($L$=16 or 32 for spatial or temporal direction,
respectively).
In the continuum limit, only $B_J^{(0)}$ and $C_J^{(1,1)}$ survive.
$B_J^{(0)}$ could contain a power divergent contribution due to a
contact term, but the exact symmetry present between the vector and
axial-vector currents guarantees that this contribution cancels
in the difference ${\Pi_V}_{\mu\nu}-{\Pi_A}_{\mu\nu}$.
Coefficients other than $B_J^{(0)}$ and $C_J^{(1,1)}$ represent lattice
artifacts.
In the difference ${\Pi_V}_{\mu\nu}-{\Pi_A}_{\mu\nu}$, these lattice
artifacts are negligible as numerically confirmed below.

\begin{figure}
 \centering
  \includegraphics*[width=0.45 \textwidth,clip=true]
  {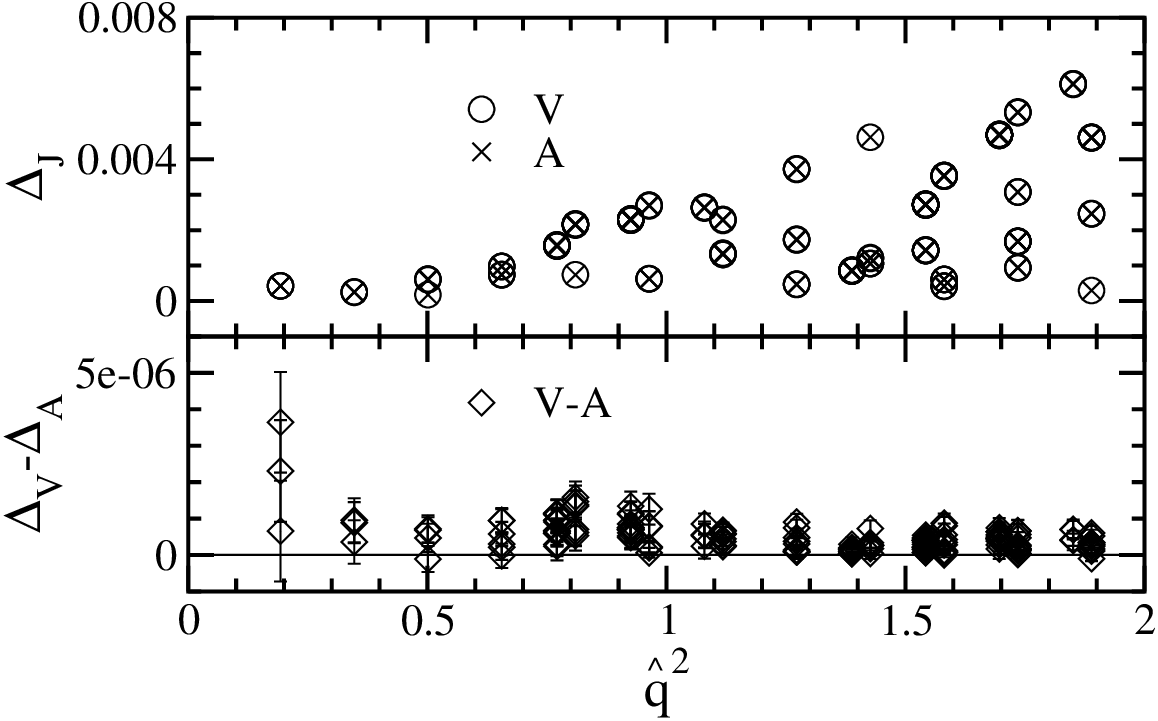}
\caption{${\hat q}^2$ dependence of $\Delta_J$ ($J=V$ or $A$) (top) and their
 difference (bottom).
 The result for $\hat m_q$=0.015 is shown.}
\label{fig:artifacts}
\end{figure}

We define a measure of the Lorentz-violating lattice artifacts by
\begin{eqnarray}
\Delta_J = \sum_{\mu,\nu}\hat q_\mu \hat q_\nu
           \left(\frac{1}{{\hat q}^2}
            - \frac{\hat q_\nu}{\sum_\lambda (\hat q_\lambda)^3} \right)
           {\Pi_J}_{\mu\nu},
\label{eq:delta_J1}
\end{eqnarray}
which contains all of $B_J^{(n)}$ and $C_J^{(n,m)}$ but $B_J^{(0)}$ nor
$C_J^{(1,1)}$.
Figure~\ref{fig:artifacts} shows $\Delta_J$ for $J=V$ and $A$ (top) and
their difference (bottom) as a function of ${\hat q}^2$ at
$\hat m_q$=0.015.
While we observe statistically significant non-zero values of $\Delta_J$
depending on $\hat q^2$, the difference is orders of magnitude smaller
than the individual $\Delta_J$.
Similar plot is obtained for $\hat m_q$=0.050.
This indicates that the Lorentz-violating lattice artifacts indeed
cancel in the difference ${\Pi_V}_{\mu\nu}-{\Pi_A}_{\mu\nu}$ and are
insensitive to S$\chi$SB or $m_q$.
Neglecting the Lorentz-violating terms, we analyze the difference
\begin{eqnarray}
{\Pi_V}_{\mu\nu}-{\Pi_A}_{\mu\nu}
= \left({\hat q}^2\delta_{\mu\nu}-\hat q_\mu \hat q_\nu\right)\Pi_{V-A}^{(1)}
  - \hat q_\mu \hat q_\nu\Pi_{V-A}^{(0)},
\label{eq:pi_v-a}
\end{eqnarray}
where $\Pi_{V-A}^{(1)}$ and $\Pi_{V-A}^{(0)}$ represent the transverse
and longitudinal contributions, respectively.

\begin{figure}
 \centering
  \includegraphics*[width=0.46 \textwidth,clip=true]
  {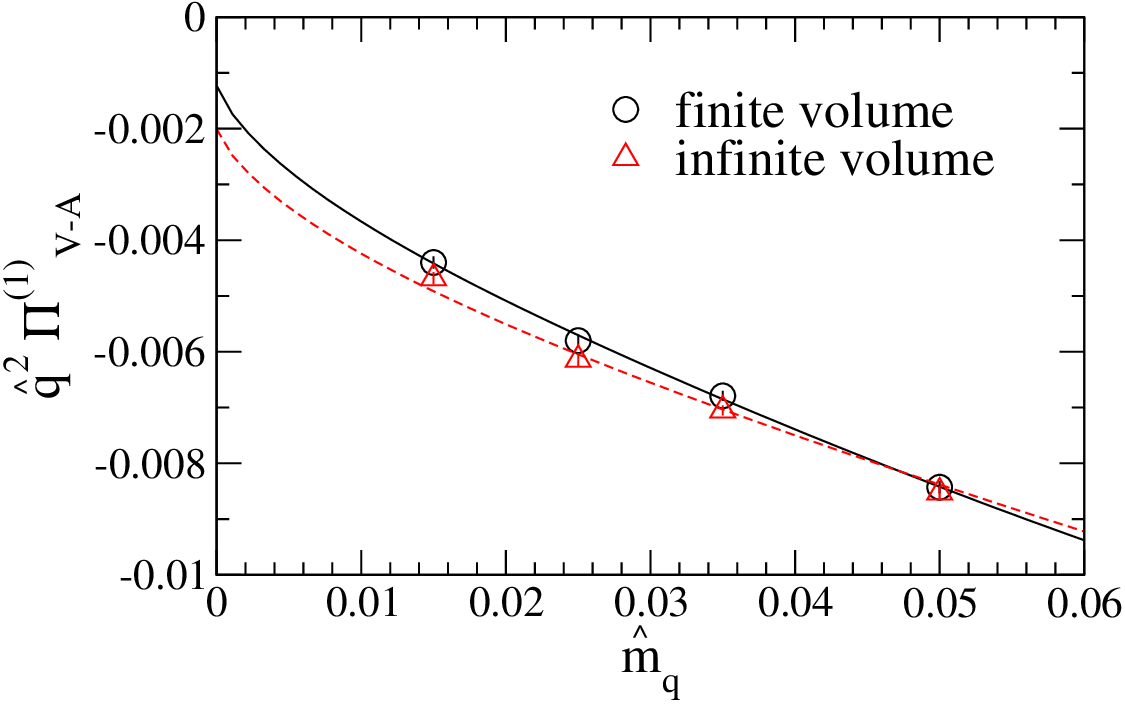}
 \caption{
 $\hat q^2\Pi_{V-A}^{(1)}|_{\hat q^2=0.038}$ as a function of quark
 mass.
 The fit results with (solid) and without (dashed) finite volume
 correction are shown.}
 \label{fig:mq-dep-pi_v-a_qmin}
\end{figure}

First we calculate $L^r_{10}(\mu)$ from $\Pi_{V-A}^{(1)}$.
At the next-to-leading order, ChPT predicts~\cite{Gasser:1983yg}
\begin{eqnarray}
&& 
    \Pi_{V-A}^{(1)}(q^2)
= - \frac{f_\pi^2}{q^2} - 8\,L_{10}^r(\mu)
  - \frac{\ln\left(\frac{m_\pi^2}{\mu^2}\right)
      +\frac{1}{3}-H(x)}{24\pi^2}, 
\label{eq:chpt_v-a_1}\\
&& 
  H(x)
= (1+x) \left[\sqrt{1+x}\,
              \ln\left(\frac{\sqrt{1+x}-1}{\sqrt{1+x}+1}
                 \right)+2\,
        \right],
\end{eqnarray}
where $x\equiv 4m_\pi^2/q^2$, and $\mu$ is a renormalization scale set
equal to the physical $\rho$ meson mass $m_\rho$.
Using the measured values of $\hat m_\pi$ and $\hat f_\pi$, we fit the
data of ${\hat q}^2\Pi_{V-A}^{(1)}$ at four quark masses
with~(\ref{eq:chpt_v-a_1}) to obtain $L_{10}^r(m_\rho)$ varying fit
range of $\hat q^2$.
Correlation among the data points are ignored since each of all the data
comes from different sea quark ensemble (also see below).
It turns out that the fit including only the smallest $\hat q^2$ point
($\hat q^2$=0.038, which corresponds to (320 MeV)$^2$) gives an
acceptable $\chi^2/$dof ($\sim$0.5).
The fit is shown in Fig.~\ref{fig:mq-dep-pi_v-a_qmin} as a function of
$\hat m_q$ (circles and solid curve).
Once the second smallest $\hat q^2$ ($\sim$ (650 MeV)$^2$ in the
physical unit) is included the fit becomes unacceptable
($\chi^2/$dof $\sim O(40)$).
This may indicate the breakdown of the chiral expansion at such a large
$q^2$.
Our result from the smallest $\hat q^2$ data is
$L_{10}^r(m_\rho)=-5.22(17)\times10^{-3}$.
Here, the error is statistical only.

We estimate the systematic error due to higher order effects of the
chiral expansion using a modified fit function to cover a wider range of
$\hat q^2$ (see below).
We obtain a slight negative shift, $0.3\times 10^{-3}$, which is added
to the systematic error.
The finite size effect may be sizable in the pion-loop effects, which is
the third term in (\ref{eq:chpt_v-a_1}), since the lattice volume (1.9
fm)$^3$ is not large enough.
We estimate its magnitude by replacing the momentum integral with a sum.
$\hat f_\pi$ and $\hat m_\pi$ are also corrected
following~\cite{Colangelo:2005gd}.
Taking these corrections into account, we fit the data at the smallest
$\hat q^2$ to~(\ref{eq:chpt_v-a_1}) and obtain
$L_{10}^r(m_\rho)|_{V=\infty}=-5.74(17)\times 10^{-3}$ with
$\chi^2$/dof= 2.3 as shown in Fig.~\ref{fig:mq-dep-pi_v-a_qmin}
(triangles and dashed curve).
We take the difference between these two results as an estimate of the
systematic errors.
We then quote
\begin{eqnarray}
 L_{10}^r(m_\rho)=-5.2(2)(^{+0}_{-3})(^{+5}_{-0})\times10^{-3},
  \label{eq:L10}
\end{eqnarray}
where the first error is statistical, and the second and third are the
estimated systematic uncertainties due to higher order effects in $q^2$
and the finite size effect, respectively.
Since only one value of $\hat q^2$ is included in the fit, the error
from the chiral fit may be underestimated.
Furthermore, other sources of uncertainty, {\it e.g.} finite lattice
spacing and lack of a dynamical strange quark, exist.
Nevertheless, (\ref{eq:L10}) is already consistent with the experimental
value $-5.09(47)\times 10^{-3}$~\cite{Ecker:2007dj}.

Next, we consider the squared-mass splitting between charged and neutral
pions.
The splitting in the chiral limit solely comes through the
electromagnetic interaction and is written by the integral of
$\hat q^2\Pi_{V-A}^{(1)}$ as given in (\ref{eq:dmpi}). 
In order to avoid possibly large discretization effects in the large
$\hat q^2$ region, we separate the whole integral region into two parts
at $\hat q^2$=2.0, and estimate each part as follows.

For the lower $\hat q^2$ region ($\le 2.0$), we fit the data to an
ansatz
\begin{eqnarray}
  \hat q^2 \Pi_{V-A}^{(1),\rm fit}(\hat q^2)
&=&-\hat f_\pi^2
 +\frac{\hat q^2 \hat f_V^2}{\hat q^2+\hat m_V^2}
 -\frac{\hat q^2 \hat f_A^2}{\hat q^2+\hat m_A^2}
 - \frac{\hat q^2}{24\pi^2}\,
     \frac{X(\hat q^2)}{1+x_5\,(Q_\rho^2)^4}
\label{eq:chpt+reson_v-a-0},
\end{eqnarray}
where $Q_\rho^2=\hat q^2/\hat m_\rho^2$ with $\hat m_\rho$ the physical
$\rho$ meson mass in lattice unit.
Here and in the following $x_i$ denotes a fit parameter.
We introduce poles of the lowest-lying state for both vector and
axial-vector channels with masses $\hat m_{V,A}$ and decay constants
$\hat f_{V,A}$.
We put the constraints
$\hat f_\pi^2 = \hat f_V^2-\hat f_A^2$ and
$\hat f_A \hat m_A = \hat f_V\hat m_V$ among them
so that they satisfy the first and second Weinberg sum
rules~\cite{Weinberg:1967kj}.
We also assume a linear dependence on $\hat m_\pi^2$:
$\hat f_V = x_1+x_3\,\hat m_\pi^2$ and
$\hat m_V = x_2+x_4\,\hat m_\pi^2$.
The function $X(\hat q^2)$ is either
\begin{eqnarray}
&&  \ln\,\left(\frac{\hat m_\pi^2}{\hat m_\rho^2}\right)
   +\frac{1}{3}-H(4\hat m_\pi^2/\hat q^2)+x_6\,Q_\rho^2,
\label{eq:chpt+reson_v-a-fit1}\\
{\rm or}&& x_6\,Q_\rho^2 \ln(Q_\rho^2).
\label{eq:chpt+reson_v-a-fit4}
\end{eqnarray}
Then, the function (\ref{eq:chpt+reson_v-a-0}) behaves as
$O(q^{-6},q^{-6}\ln q^2)$ at large $q^2$ in the chiral limit, which is
consistent with the asymptotic scaling predicted by the operator product
expansion (OPE)~\cite{Braaten:1991qm}.
Taking (\ref{eq:chpt+reson_v-a-fit1}) for $X(\hat q^2)$,
$\Pi_{V-A}^{(1),\rm fit}(\hat q^2)$ reduces to the ChPT prediction
(\ref{eq:chpt_v-a_1}) when $Q^2_\rho \ll 1$,
while (\ref{eq:chpt+reson_v-a-fit4}) gives a logarithmic term in the
large $Q_\rho^2$ region as expected by OPE.

\begin{figure}
 \centering
  \includegraphics*[width=0.46 \textwidth,clip=true]
  {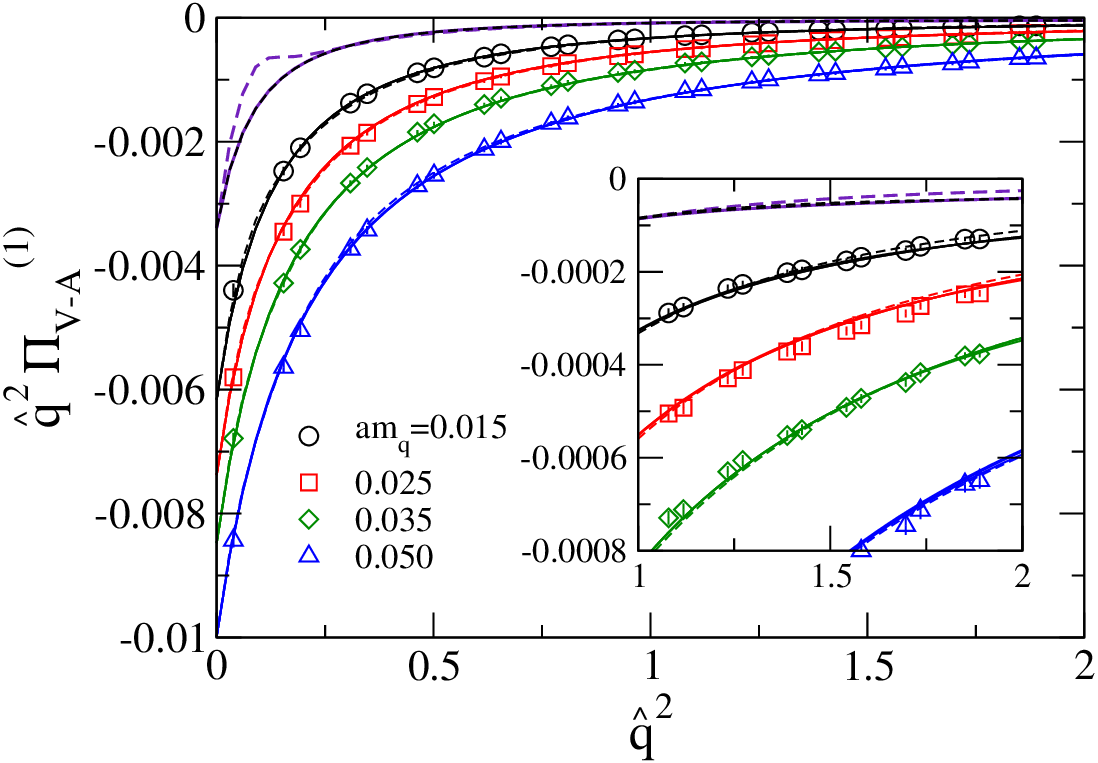}
 \caption{The fit results with (\ref{eq:chpt+reson_v-a-fit1})
 (dashed curves) and (\ref{eq:chpt+reson_v-a-fit4}) (solid curves).
 The results in the chiral limit are also shown.
 The statistical errors shown are smaller than the symbol size.}
 \label{fig:result}
\end{figure}

We fit the data at $\hat q^2 \le 2.0$ using the measured values of
$\hat f_\pi$ and $\hat m_\pi$ as shown in Fig.~\ref{fig:result}.
We have only attempted an uncorrelated fit since the full covariance
matrix is likely ill-determined for many data points and free parameters
in this fit.
Both (\ref{eq:chpt+reson_v-a-fit1}) and (\ref{eq:chpt+reson_v-a-fit4})
fit the data quite well and indeed give a reasonable $\chi^2$/dof,
though the latter is slightly better.
Integrating over $\hat q^2$ in the chiral limit, we obtain
$m_{\pi^\pm}^2|_{\hat q^2\le 2.0}$= 676(50) and 811(12) MeV$^2$ for
(\ref{eq:chpt+reson_v-a-fit1}) and (\ref{eq:chpt+reson_v-a-fit4}),
respectively.
The difference arises from the chiral extrapolation around
$\hat q^2$=0.1--0.2, since (\ref{eq:chpt+reson_v-a-fit1}) contains the
chiral logarithmic term.
Recalling that in the determination of $L_{10}^r$ the ChPT formula fits
the data only at the smallest $\hat q^2$ and
(\ref{eq:chpt+reson_v-a-fit4}) fits the data better than
(\ref{eq:chpt+reson_v-a-fit1}), we take the central value from the fit
with  (\ref{eq:chpt+reson_v-a-fit4}) and the difference as a systematic
error due to the chiral extrapolation.

Expanding $\hat q^2\Pi_{V-A}^{(1),\rm fit}$ around $\hat q^2=0$ in the
chiral limit and comparing with (\ref{eq:chpt_v-a_1}),
we obtain $L_{10}^r(m_\rho)$=
$-\hat f^2\,(2 x_1^2-\hat f^2)/(8\,x_1^2 x_2^2)$.
With the fit results for (\ref{eq:chpt+reson_v-a-fit1}),
this gives $L_{10}^r(m_\rho)=-4.9\times 10^{-3}$.
The difference from the central value is added to the systematic error
from the higher order effect, and included in (\ref{eq:L10}) as
already mentioned.

The remaining part of the integral ($\hat q^2\ge$ 2.0) is estimated
based on the OPE, which predicts
$\Pi^{(1)}_{V-A}(q^2) \sim a_6/(q^2)^3$ in the chiral limit for large
$q^2$ up to a logarithmic term.
Assuming $\Pi^{(1)}_{V-A}|_{\hat m_q=0} = a_6/(\hat q^2)^3$ at
$\hat q^2$=2, the fit result with (\ref{eq:chpt+reson_v-a-fit4}) gives
$a_6$=$-$0.0035.
In the estimate of the final result, we use a phenomenological value in
the range [$-$0.001,\ $-$0.01] GeV$^6$~\cite{Bijnens:2001ps} to be
conservative.
An integral then gives
$m_{\pi^\pm}^2|_{q^2\ge 2.0}$= 182(149) MeV$^2$. 

Summing up the two parts, we obtain
\begin{equation}
 m_{\pi^\pm}^2= 993(12)(^{+\ \ 0}_{-135})(149)\ \mbox{MeV$^2$},
\end{equation}
as the pion squared-mass splitting in the chiral limit.
The first error is statistical; the second and third ones are due to the
chiral extrapolation and the uncertainty in $a_6$.
The result is reasonably consistent with the experimental value at the
physical quark mass [1261 MeV$^2$].
In addition to the errors quantified above, other sources of systematic
errors may still remain.
We do not expect, however, substantial systematic errors other than
those estimated above, since the integral is dominated by the
$\hat q^2\sim$ 0 region where the integrand
$\hat q^2\Pi_{V-A}^{(1)}/\hat f^2$ in the chiral limit is strongly
constrained by the first Weinberg sum rule
$[\hat q^2\Pi^{(1)}_{V-A}]_{\hat q^2=0}/\hat f^2$=1.

In this letter, we have demonstrated that the $S$-parameter and the pNGB
mass can be calculated using the lattice QCD technique.
Since these quantities are generated solely through S$\chi$SB, the exact
chiral symmetry on the lattice plays an essential role to prohibit
contaminations from the explicit breaking.
The method is general and the application to other vector-like gauge
theories with arbitrary number of colors and flavors is straightforward.
Thus with this method the lattice technique is able to directly
investigate physical quantities relevant for the LHC phenomenology.
In addition to these quantities, we can also calculate $a_6$ and the
strong coupling constant using the data in the large $q^2$ region.
The results will be reported in a subsequent paper.

We thank M. Golterman for useful comments.
The work of HF is supported by Nishina Memorial Foundation.
This work is supported in part by the Grant-in-Aid of the Japanese
Ministry of Education
(Nos.
      17740171, 
      18034011, 
      18340075, 
      18740167, 
      19540286, 
      19740121, 
      19740160, 
      20025010, 
      20340047  
).
Numerical simulations are performed on Hitachi SR11000 and IBM System Blue
Gene Solution at High Energy Accelerator Research Organization (KEK) under a
support of its Large Scale Simulation Program (No.~07-16).

\bibliography{basename of .bib file}

\end{document}